# sI clathrate hydrate confined in $C_8$ grafted SBA-15: A high-efficiency methane storage system enabling ultrafast loading and unloading

Emile Jules Beckwée[a], Maarten Houlleberghs[b], Radu-George Ciocarlan[c], C. Vinod Chandran[b], Sambhu Radhakrishnan[b], Lucas Hanssens[b], Pegie Cool[c], Johan Martens[b], Eric Breynaert[b], Gino V. Baron[a], Joeri F. M. Denayer[a,*]

[a]Department of Chemical Engineering, Vrije Universiteit Brussel, Pleinlaan 2, 1050 Brussel, Belgium

[b]Centre for Surface Chemistry and Catalysis, NMRCoRe - NMR - XRAY - EM Platform for Convergence Research, Department of Microbial and Molecular systems (M2S), KU Leuven, Celestijnenlaan 200F, 3001 Leuven, Belgium

[c]Laboratory of Adsorption and Catalysis, Department of Chemistry, University of Antwerp, Universiteitsplein 1, 2610 Wilrijk, Belgium

E. J. Beckwée and M. Houlleberghs contributed equally to this work.

*Corresponding author: Joeri F.M. Denayer (E-mail: joeri.denayer@vub.be)

# Abstract

Dispersion and confinement of water in the mesopores of a hydrophobized SBA-15 material was found to promote methane clathrate hydrate nucleation and growth, significantly enhancing formation kinetics. Methane clathrate formation was studied in as synthesized SBA-15, as well as in SBA-15 grafted with hydrophobic octyl groups (SBA-15 $C_8$) to render the mesopores more hydrophobic. The water loading was optimized at 150% of the pore volume of the host material ($R_w$ = 150%). High-pressure methane uptake isotherms revealed a 71 % water-to-hydrate conversion for SBA-15, whereas almost the entire mass of water (96 %, 14.8 g $CH_4$/g $H_2O$) was converted using the hydrophobized SBA-15 $C_8$ material at 7.0 MPa and 248 K. X-ray diffraction demonstrated formation of structure I methane hydrate in the presence of SBA-15 $C_8$ particles. Using 2D NMR correlation spectroscopy ($^1H$-$^{13}C$ CP-HETCOR, $^1H$-$^1H$ RFDR) clathrate hydrate formation was shown to not only occur in interstitial volumes, but also within the pores of SBA-15 $C_8$. Following the initial formation of SBA-15 $C_8$-supported methane clathrate hydrate ($t_{90}$ = 207 min), pressure-induced hydrate dissociation at 248 K and subsequent reformation was achieved within minutes ($t_{90}$ = 7.8 min), at half the energy cost compared to thermal degradation of the clathrate hydrate at 298 K and subsequent reformation at 248 K. Such fast nucleation and growth persisted throughout 19 subsequent dissociation-reformation cycles. $^{13}C$ and $^2H$ NMR analysis of a dissociated $D_2O$ hydrate supported by SBA-15 $C_8$ revealed the presence of a small amount of residual methane hydrate in combination with an amorphous yet locally ordered ice phase, which was absent prior to the initial hydrate formation. This molecular-scale evidence assists to explain the enhanced reformation kinetics observed at the macroscopic level, revealing new perspectives for hydrate-based methane storage.

Keywords: SBA-15, methane hydrate, clathrate hydrate, NMR, ice, formation kinetics

# 1 Introduction

Methane clathrate hydrates (abbr. methane hydrates) are crystalline molecular solids built up from cages of hydrogen-bonded water molecules containing a guest molecule, i.e., methane, stabilizing the structure through van der Waals interactions.[1] In nature, immense deposits of methane hydrates are present in every location where water and methane occur together in an environment with a suitable pressure/temperature regime, i.e. in deep-sea sediments and in permafrost regions.[2,3] These natural methane hydrate resources are estimated to contain up to $1.8 \cdot 10^{15}$ kg of methane[4]. Synthetic methane hydrates, obtained by contacting water and methane under high-pressure (P ~ 2.5 – 11 MPa) and at low-temperature (T ~ 273 – 285 K)[5], bear great promise as storage medium.[6] Among the different clathrate hydrate structures, especially structure I (sI) hydrates have gained significant scientific interest because of their superior methane storage capacity.[7] The cubic unit cell of an sI type hydrate comprises 2 small cages and 6 large cages of hydrogen-bonded water molecules. In the small cages, these water molecules reorganize into a polygon consisting of 12 pentagonal faces ($5^{12}$), whereas the large cages are composed of 12 pentagonal faces and 2 hexagonal faces ($5^{12}6^2$).[1] Each of these cages can accommodate one methane molecule, resulting in stoichiometry of 8 $CH_4$:46 $H_2O$ per unit cell. This corresponds to a maximum gravimetric storage capacity of 15.5 wt% methane, equivalent to 172 v/v (volume of gas at STP/volume of hydrate),[8] substantiating the practical relevance of methane hydrate technology as a cost-effective and safe alternative for (large-scale) methane storage compared to liquefaction and compression, i.e., LNG (T ≤ 191 K) and CNG (P ~ 20 – 25 MPa), respectively.[6]

Hydrate-based methane storage has been hampered by the intrinsically slow formation rate of the hydrate phase, especially in bulk water systems.[9] Following the initial formation of a hydrate thin film at the methane-water interface, hydrate growth is governed by diffusion of the gaseous guest molecules to the hydrate-water interface, across the hydrate film. As the film grows, methane diffusion drastically slows down, decreasing hydrate formation rate by up to two orders of magnitude, and is ultimately halted once a hydrate film thickness of 20 – 100 μm is reached (depending on the experimental conditions).[10,11] The sluggish kinetics have in part been overcome by adding kinetic promotors to the mix such as surfactants, polymers, amino acids, *etc*. to enhance the concentration of otherwise poorly water-soluble compounds, like methane, speeding up hydrate formation and increasing the overall water-to-hydrate conversion.[12] Addition of hydrocarbons such as neohexane and cyclopentane, and heterocyclic compounds including tetrahydrofuran (THF) and dioxane has also been shown to promote hydrate formation. Rather than affecting kinetics, these compounds mostly impact the thermodynamic stability of the hydrate system shifting its phase equilibrium to higher temperatures and lower pressure.[8,12–14] This beneficial effect is believed to originate from the local reorganization of the hydrogen-bonded water structure surrounding the solvated promotor molecules, i.e., their hydration shell, facilitating methane enclathration.[15] In the case of 5.6 mol% THF as promoter, Veluswamy et al. reported rapid methane hydrate formation at 283.2 K and 7.2 MPa, showcasing both kinetic and thermodynamic promotion.[16] Unfortunately, addition of thermodynamic promoters comes at the expense of methane storage capacity as the promoter typically occupies the larger clathrate cages.[12]

More recently, porous materials have been put forward as an alternative for promoting methane hydrate formation without requiring any of the aforementioned additives.[17,18] Irregularities on the surface of the solid host impact the local water structure, creating heterogeneous nucleation sites which will eventually lead to an amorphous precursor phase promoting hydrate crystallization.[19] Also, the intrinsic porosity of the host enables to efficiently disperse the water and maximize the gas-water interface, enhancing the kinetics of formation, while the spatial confinement of the hydrate phase in the pore is hypothesized to provide additional support, cfr. the quasi-high-pressure effect typically observed in nanopores,[20–22] stabilizing the hydrate structure in otherwise thermodynamically unstable temperature and pressure conditions.[23–25] Using a carbon-based material as porous host, Casco et al. reported the fast (within minutes) and fully reversible synthesis of an sI methane hydrate at 275 K and 3.5 MPa, containing 15.4 wt% methane with respect to the aqueous phase.[23] Aside from carbons[26–34], other types of porous materials have also been shown to positively impact hydrate formation kinetics and thermodynamics, including metal-organic frameworks[35–40], zeolites[41–44], polymers[45,46], and silica-based materials.[47–65] The extent to which a certain porous material can promote hydrate formation is believed to be largely determined by its pore size and surface chemistry.[66] In a systematic study on methane uptake by wet micro-, meso- and macro-porous carbons, Borchardt et al. showed that mesopores with a pore size of 25 nm are superior in both storage capacity and nucleation and growth rate at 264 K and 8.5 MPa methane.[29] Micropores (< 2 nm) are generally thought of to be too small to host sI clathrate hydrate structures (unit cell lattice: 1.2 nm), resulting in slow formation kinetics, while the volume-to-surface ratio of the macropores is too large to impose the desired confinement effect.[29,49] The surface chemistry of the pores can either be hydrophilic or hydrophobic in nature. Hydrophilic materials have the advantage of better water adsorption, facilitating the dispersion of water throughout the pore network. Oversaturation of the pores, however, can become an issue and can lead to exclusion of methane from the pore volume, slowing down hydrate formation as it will primarily take place on the external surface.[66,67] Also, hydrogen bond formation between water and polar moieties in the pore wall, i.e. oxygen or nitrogen atoms, decreases the water activity, shifting the thermodynamic threshold for methane hydrate formation to higher pressures at a given temperature or to lower temperatures at a given pressure.[17,66] This thermodynamic inhibition effect is absent in porous materials with hydrophobic surfaces, as the strength of the water-surface interactions becomes negligible compared to the water-water and water-methane interactions, stabilizing the hydrate structure.[30] Consequently, water molecules tend to cluster in the center of the pore, away from the hydrophobic surface, allowing methane to enter the pores and accumulate in a thin layer between the pore surface and the water clusters.[37] The increased gas density, along with the increased water-gas contact area, is believed to play a key role in the promotion of gas hydrate formation near hydrophobic surfaces.[68,69] In addition, water clustering imposes a higher degree of ordering on the hydrogen bonding network, which should facilitate its phase transition to ordered clathrate hydrate structures when contacted with methane at low temperatures as compared to more hydrophilic surfaces.[30,56,70] This was recently demonstrated by Casco *et al.*, who studied methane hydrate formation in MCM-41 (hydrophilic) and B-PMO materials (hydrophobic).[49] At 243 K and 6 MPa, the B-PMO sample displayed shorter induction times and higher water-to-hydrate conversion ratios.[49] The main disadvantage of purely hydrophobic materials, however, lies in the high pressures required for the water to penetrate the pores, cfr. the Washburn's equation.[71] Up to 15 MPa is needed to intrude water into 10 nm cylindrical pores (assuming a contact angle of 120 °), for instance, and this pressure continues to increase with decreasing pore size.[72]

Literature reports on hydrate formation in porous materials often describe the porous host as either hydrophilic or hydrophobic. The ideal porous host, however, is expected to combine the best of both worlds, facilitating water uptake and dispersion in its (meso)pores while still enabling (local) modulation of the water structure.[73] This work documents the preparation and application of an SBA-15 material grafted with trimethoxy(octyl)silane. Selection of n-octyl groups over other alkyl functionalities was based on previous exploratory work. The grafting procedure relies on the hydrolysis of the alkoxy moieties and condensation of the resulting alcohol with surface silanols, which are concentrated in the pores given the large internal surface area of the material.[74] Introduction of n-octyl groups ($C_8$) onto the SBA-15 surface (SBA-15 $C_8$) comes at the expense of silanols, thus hydrophobizing the surface of the porous host. $C_8$ grafting was selected based on a preliminary *in situ* synchrotron X-ray diffraction study performed at the DUBBLE beamline (ESRF, Grenoble) using a commercial $C_8$ grafted reverse phase silica[75,76]. The performance of the modified material as methane hydrate promotor is compared to that of a pristine SBA-15 sample, and the optimal water loading of the SBA-15 $C_8$ material is determined in a series of gravimetric uptake measurements. Powder X-ray diffraction and solid-state nuclear magnetic resonance spectroscopy are used to characterize the hydrate phase. Reversible emptying and filling of the supported hydrate phase in consecutive pressure-swing cycles was investigated with respect to the development of a practical methane storage process.

## 2 Materials & methods

### 2.1 Sample synthesis

Mesoporous silica SBA-15 was obtained following a well-known procedure in literature.[77] In the procedure, the surfactant triblock copolymer P123 ($EO_{20}PO_{70}EO_{20}$) was dissolved in a mixture of water and HCl. Next, tetraethyl orthosilicate (TEOS) was added, as Si-source, and the white precipitate was stirred for 7.5 h at 318 K, followed by a static aging overnight (15.5 h) at 353 K. Finally, the mixture is washed with distilled water, dried at 333 K and calcined at 823 K for 6 h with a ramp of 1 K min$^{-1}$. The molar ratios of the reactants were 1 TEOS: 5.87 HCl: 194 $H_2O$:0.017 P123. Note that, for the non-functionalized SBA-15 sample, the same procedure was followed, but to reduce the pore size diameter, the aging was done in an autoclave reactor vessel.

Post synthesis surface modification followed a slightly modified procedure from literature.[78] Firstly, 3 g of SBA-15 was dried for 3 h at 473 K and suspended in 100 mL of toluene, in a glovebox setup. When the mixture became homogeneous, 10 g of trimethoxy(octyl)silane was added and the translucent solution was stirred for 3 days at 353 K. Finally, the precipitate was washed 6 times with 50 mL toluene to remove the unattached/unreacted surface modifier, followed by an overnight drying process at 323 K. The sample was denoted SBA-15 $C_8$, correlated with the grafted linear aliphatic chain of the modifier.

### 2.2 Sample characterization

The micro- and mesoporosity of the samples was evaluated by measuring argon (Ar) physisorption at 87 K on a volumetric Autosorb 1 device. Before measuring, samples were activated for 12 h at 393 K under vacuum, reaching this temperature at 2 K min$^{-1}$. The pore size distribution, pore volume and surface area were determined by applying non-local density functional theory (NLDFT) to the adsorption branches of the isotherms (NLDFT method: Ar at 87 K - zeolite/silica - spherical/cylindrical pores). Water adsorption at 303 K was measured gravimetrically under flow on a VTI Corporation SGA-100H instrument using nitrogen as carrier gas. Before measuring, samples were activated under 200 Nml min$^{-1}$ nitrogen by heating at 2 K min$^{-1}$ to 393 K and maintaining at this temperature for 4 h. Functional group density was determined by combining thermogravimetric analysis (TGA), measured on a Mettler Toledo TGA/DSC 3+ Star system, and the Ar physisorption derived Brunauer–Emmett–Teller (BET) surface area (Eq. 1). For the equation below, the mass loss considered was only after 473 K until 1073 K. The mass loss below this temperature is attributed to residual solvents used in the rinsing step.

$$\rho_{C_8} = \frac{\Delta m \; N_A}{M_{C_8} \; SA} \tag{1}$$

With $\rho_{C_8}$ the functional group density (groups nm$^{-2}$), $\Delta m$ the mass loss during TGA expressed per 100 g of sample (7.0 wt%), $M_{C_8}$ the molar mass of the functional group (113 g mol$^{-1}$), SA the BET surface area of the SBA-15 before modification (m$^2$ g$^{-1}$) and $N_A$ the Avogadro constant (6.022 10$^{23}$ mol$^{-1}$). A Nicolet 6700 Fourier Transform IR spectrometer was used to perform in-situ diffuse reflectance infrared FT (DRIFT) measurements. The samples were heated to 393 K, under vacuum, for 20 minutes and this temperature was maintained during the spectra acquisitions. The samples were diluted in KBr (2 wt%), 100 scans were accumulated for each spectrum with a resolution of 4 cm$^{-1}$ in the 4000–500 cm$^{-1}$.

## 2.3  High-pressure methane (hydrate) uptake

High-pressure methane uptake was measured on an in-house developed gravimetric device, based on a Rubotherm magnetic suspension balance coupled to an in-house developed gas dosing system (Fig. S1). A Keller pressure transducer ensures quantification between vacuum and 15 MPa with an accuracy of 0.05 %, and a Julabo Dyneo DD-1000F circulation thermostat ensures temperature quantification and control between 243 K and 343 K within 0.01 K. For physisorption measurements on dry adsorbent, samples were activated in-situ by increasing the temperature to 393 K at 2 K min$^{-1}$, and maintaining at 393 K for 12 h under vacuum. For methane hydrate measurements, samples were activated ex-situ at 393 K for 12 h under vacuum. Deionized water (18.2 MΩ cm resistivity) was added dropwise to the activated samples until the desired water loading was achieved. Water loading is expressed as a volume percentage relative to the total pore volume derived from Ar physisorption at 87 K (Eq. 2).

$$R_W = \frac{m_{H_2O}}{m_A \; \rho_{H_2O} \; V_{tot}} * 100 \; \% \tag{2}$$

With $R_w$ the water loading relative to the pore volume (%), $m_{H_2O}$ the mass of water added to the activated sample (g), $m_A$ the mass of activated adsorbent (g), $\rho_{H_2O}$ the density of water (1 g cm$^{-3}$) and $V_{tot}$ the total micro- and mesopore volume derived from argon adsorption at 87 K (cm$^3$ g$^{-1}$).

After activation and sample preparation, however prior to methane uptake measurements, a buoyancy experiment was performed to determine the mass and volume of the sample and its container and hence take the effect of buoyancy into account during uptake measurements. Briefly, a non-adsorbing gas, i.e. helium (He), was incrementally dosed to the sample cell. At each pressure step an equilibration time of 30 min was adopted, during which the sample mass and gas density were recorded. For dry ($V_w = 0.0$ %) samples, the experiment was conducted at 303 K, while for water supplied samples, the experiment was conducted at 248 K to prevent water loss through evaporation or sublimation at vacuum. Linear regression of the mass versus density plot allows to determine the sample mass and volume. An equivalent experiment without sample allows to determine the mass and volume of the sample container.

High-pressure methane uptake/release experiments were conducted at 248 K by dosing/evacuating methane from the sample cell while recording mass changes. Exact knowledge of the sample mass and volume allows to correct mass measurements for the buoyancy force and hence allows to determine the specific amount adsorbed or total storage capacity (Eq. 3).

$$q_{CH_4}^T = \frac{m_{CH_4}}{m_{H_2O} + m_A} * 100\ \% \tag{3}$$

With $q_{CH_4}^T$ the total storage capacity (wt%), $m_{CH_4}$ the surface excess methane uptake (g) and $m_{H_2O}$ the mass of water added to the adsorbent (g).

The water loading allows to express methane uptake in terms of water content (Eq. 4) or mass of dry adsorbent (Eq. 5).

$$q_{CH_4}^W = q_{CH_4}^T \left(1 + \frac{m_A}{m_{H_2O}}\right) * 100\% \tag{4}$$

$$q_{CH_4}^A = q_{CH_4}^T \left(1 + \frac{m_{H_2O}}{m_A}\right) * 100\% \tag{5}$$

With $q_{CH_4}^W$ the hydrate storage capacity (wt%) and $q_{CH_4}^A$ the dry weight storage capacity (wt%). Furthermore, $q_{CH_4}^W$ allows to calculate the water-to-hydrate conversion:

$$Y = \frac{q_{CH_4}^W}{q_{CH_4}^{W,sI}} * 100\% \tag{6}$$

With Y the water-to-structure I methane hydrate conversion (%) and $q_{CH_4}^{W,sI}$ the hydrate storage capacity of structure I methane hydrate (1 $CH_4 \cdot$ 7.75 $H_2O$ stoichiometry) (15.5 wt%).

## 2.4 NMR spectroscopy

NMR measurements were performed on SBA-15 $C_8$ $R_w$ = 150 %. $D_2O$ was preferred over $H_2O$ as it allows to more easily discriminate between protons originating from $CH_4$, residual $H_2O$ and host material. The sample was packed in a 7 mm NMR rotor with open $ZrO_2$ cap, placed in a tailor-made Swagelok® setup and cooled down to 248 K at 7 MPa. $^{13}C$ enriched methane gas ($^{13}CH_4$, Sigma-Aldrich, 99 % enrichment) was used to pressurize the sample. After hydrate formation, the rotor was cooled with liquid nitrogen and transferred to a wide-bore 500 MHz Bruker NMR spectrometer equipped with a HXY 7mm triple resonance MAS probe pre-cooled to 223 K.

The 2D correlation NMR spectroscopy was performed on samples loaded with $^{13}C$-labelled methane. $^1H$-$^1H$ RFDR and $^1H$-$^{13}C$ CP-HETCOR experiments were run at a MAS frequency of 4.5 kHz and a temperature of 223 K. The 2D RFDR experiment used RF pulses of strength 80 kHz with a recycle delay of 2 s and 32 scans were collected for each slice. With an increment of 10 ms, 1024 slices were recorded. The 2D CP-HETCOR experiment was carried out using a CP contact time of 8 ms. A recycle delay of 2 s was used for 640 scans per slice of the 2D experiment. With an increment of 111 ms, 50 slices were recorded. Effective $^1H$-decoupling was done using $SW_f$-SPINAL sequence during the acquisition.[79] The chemical shifts were referenced against the secondary standard adamantane.

Solid-state $^2H$ NMR experiments were done under static condition in a HXY 7mm triple resonance MAS probe in Bruker Avance III 500 MHz NMR spectrometer, operating at 11.4 T static magnetic field with a $^2H$ Larmor frequency of 76.89 MHz. Standard solid-echo (quadrupolar echo) sequence was used to collect the $^2H$ NMR spectra using RF pulses of strength 40 kHz with a recycle delay of 2 s. 2048 number of transients were recorded for each spectrum. The experiments were conducted at 223 K using an external cooling unit with liquid nitrogen. Spectral decomposition was done using DMfit program.[80]

## 2.5 X-ray diffraction

The crystal structure of methane hydrate formed using SBA-15 $C_8$ $R_w$ = 150 % was characterized by X-ray diffraction. To account for reflections originating from crystallinity of the mesoporous silica host itself, a reference diffraction pattern of SBA-15 $C_8$ was recorded prior to hydration and pressurization. To ensure stability of the hydrate phase at atmospheric pressure, the original Mylar® film in an existing XRD sample holder was replaced by Kapton® to enable operation at liquid nitrogen ($LN_2$) temperature, i.e., 77 K. Methane hydrate containing samples formed at 248 K and 7.0 MPa were first ground in a mortar filled with $LN_2$ and then transferred to the sample holder which had been submerged in $LN_2$ up until the transfer. PXRD data were recorded using a B STOE STADI P Combi diffractometer with focusing Ge(111) monochromator (CuK$_{\alpha 1}$ radiation, λ = 0.154 nm) in a high throughput set-up in transmission geometry. Data was collected using a 140°-curved image plate position sensitive detector (IP PSD) from –17.5 to 62.5° 2Θ.

## 3 Results and discussion

### 3.1 SBA-15 synthesis and characterization

In Section 3.2, results of methane hydrate formation promoted by SBA-15 $C_8$ will be compared to hydrate formation on a non-functionalized SBA-15. In the presence of a porous solid, hydrate nucleation and growth conditions are heavily influenced by pore size[81,82] and surface chemistry.[30,35,49,83] Hence, rather than using the parent/nongrafted material of SBA-15 $C_8$, which inevitably has a larger average pore size than its functionalized counterpart, it was chosen to use an SBA-15 with a more similar pore size distribution to the present SBA-15 $C_8$. Both samples were synthesized following the procedure described in Section 2.1 of Materials & methods. The presence of $C_8$ groups, chemically bonded to the surface of SBA-15, was confirmed by *in situ* FTIR analyses. More precisely, in Fig. S2, the very sharp characteristic band of isolated, free silanols can clearly be observed around 3745 cm$^{-1}$ for SBA-15, while this band is drastically smaller for SBA-15 $C_8$. Thermogravimetric analysis showed almost 7 % carbon content present in the SBA-15 $C_8$ belonging to the modifying agent (Fig. S3). Textural characterization by Argon (Ar) physisorption at 87 K shows a similar total pore volume (SBA-15: 0.715 cm$^3$ g$^{-1}$, SBA-15 $C_8$: 0.649 cm$^3$ g$^{-1}$) and surface area (SBA-15: 438 m$^2$ g$^{-1}$, SBA-15 $C_8$: 452 m$^2$ g$^{-1}$) for both materials. Applying non-local density functional theory (NLDFT) to the adsorption branches, allows to retrieve pore size distributions (Fig. S4b). Both samples exhibit quite narrow mesopore distributions centered around 6.7 nm. A detailed description of the porosimetry and an overview of pore volumes is provided in Section S4. Combining the thermogravimetric result and the Ar-physisorption derived surface area, a functional group density of 0.6 groups nm$^{-2}$ can be calculated for SBA-15 $C_8$ (Eq. 1). Surface tailoring by alkyl groups at given density clearly alters the hydrophobicity, as demonstrated by water vapour adsorption isotherms, showing an upshifted capillary condensation point for the alkyl grafted material (SBA-15: p/p$_0$ = 0.60, SBA-15 $C_8$: p/p$_0$ = 0.80) (Fig. S5).

## 3.2 CH$_4$ hydrate formation on SBA-15 and SBA-15 C$_8$

First, the influence of pressure on methane hydrate formation promoted by SBA-15 C$_8$ was assessed and compared to hydrate formation on a non-functionalized SBA-15. High-pressure gravimetric methane hydrate isotherms were measured on both materials at 248 K (**Error! Reference source not found.**a). Both samples were loaded with a volume percentage of water to total pore volume (R$_w$) of 150 % (Eq. 2). In Fig. 1, methane uptake is normalized to the water loading of samples, i.e., the hydrate storage capacity $q_{CH_4}^w$ is displayed. Capacities may easily be converted relative to the solid or total mass by Eq. 4 and 5.

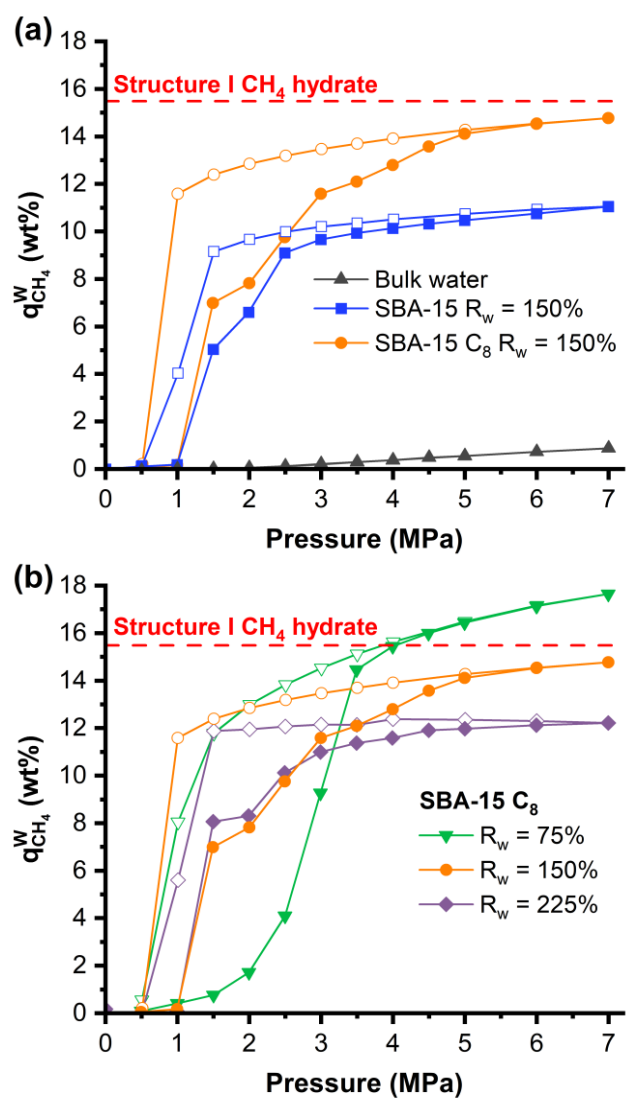

**Fig. 1** High-pressure gravimetric methane hydrate isotherms measured at 248 K on (a) SBA-15 (R$_w$ = 150 %), SBA-15 C$_8$ (R$_w$ = 150 %) and bulk water, and (b) on SBA-15 C$_8$ at different water loadings. Formation and dissociation points are shown by filled and hollow markers, respectively. The theoretical gravimetric methane capacity of structure I hydrate is shown by the horizontal dashed red line.

First, under wetted conditions both samples exhibit isotherm shapes completely dissimilar to those measured under dry conditions ($R_w$ = 0 %) (Fig. S6), indicating a different or complementary methane uptake mechanism besides adsorption, i.e. methane hydrate formation. The isotherms are in accordance with what could be called a three-regime methane uptake mechanism, previously reported for methane hydrate formation aided by mesoporous silica[49] and other solid promotors.[23] i) At low pressure (0 – 1.0 MPa) population of water on pore surfaces and in between solid particles leads to a blocking effect and thereby reduced physisorption compared to the dry sample ($R_w$ = 0 %). ii) Around the phase boundary of natural methane hydrate, that is 1.1 MPa at 248 K,[84] a clear increase in methane uptake is observed, commonly attributed to hydrate formation within large pores or interstitial spaces. iii) At higher pressures (2.0 – 7.0 MPa) the isotherm exhibits a more gradual increase, ascribed to hydrate formation within mesopores and large micropores. Although similar shaped isotherms are obtained for both samples, SBA-15 clearly shows inferior methane uptake over the assessed pressure range. At the initial hydrate formation pressure, that is 1.5 MPa, 33 % of water is converted to hydrate for SBA-15, while 45 % for SBA-15 $C_8$. At 7.0 MPa, the highest measured pressure, almost the entire mass of water is converted to hydrate for the functionalized material (96 %), while only 71 % for the other. This result is in line with the promoted gas hydrate formation near a hydrophobic surface, discussed in section 1. Zhou *et al.* measured methane hydrate isotherms on SBA-15 $R_w$ = 150 % as well, and reported a similar water-to-hydrate conversion of 32 % at 3.9 MPa and 275 K. However, rather than an initial jump in capacity followed by a more gradual increase, a single stepped isotherm shape was observed.[65] Up to 10 MPa, no clear increase in methane uptake was measured, potentially explained by the need for even higher pressures to convert the remaining water to hydrate at this temperature.

Although the positive effect of surface hydrophobicity on methane hydrate formation has well been recognized by multiple authors[30,35,49,57,68,85], a comparison between a mesoporous silica deliberately grafted with a single type of functional group and an ungrafted counterpart allows for a unique, more direct and undistorted evaluation of the hydrate promoting nature of a solid since both have same pore geometries and connectivities and very similar pore sizes and volumes.

The desorption branches of both samples show progressive reduction of methane capacity with decreasing pressure and a sudden drop in the low-pressure region, most likely associated to complete dissociation of the hydrate phase. Interestingly, while for the non-grafted sample, majority of methane is released at 1.0 MPa, that is when the methane hydrate phase boundary is crossed[84], a further decrease to 0.5 MPa is required for the modified sample. It thus appears that for SBA-15 $C_8$ the hydrate phase remained stable outside of natural hydrate equilibrium conditions, at least during the adopted 4.7 h equilibration time, further demonstrating the positive effect of hydrophobizing the surface on methane hydrate stability. Finally, as a frame of reference, methane uptake on bulk water was measured (**Error! Reference source not found.**a). The isotherm shows very little methane uptake over the whole pressure range (e.g. $q_{CH_4}^w$ = 0.9 wt% or 6 % water-to-hydrate conversion at 7.0 MPa), effectively proving the promoting effect of confining and distributing water by either of both porous solids on hydrate nucleation and growth.

Since the hydrophobized SBA-15 showed superior methane hydrate formation, SBA-15 $C_8$ was further studied by measuring methane hydrate isotherms at varying water loading ($R_w$ = 75 %, 150 %, 225 % and 300 %) (Fig. 1b and S7). The isotherm of the undersaturated sample ($R_w$ = 75 %) shows a sigmoidal shape, without a clear onset of hydrate formation. Although methane uptake is shifted to higher pressures compared to the oversaturated samples ($R_w$ = 150 % and $R_w$ = 225 %), the capacity in the high-pressure region is the highest of all water loadings ($R_w$ = 75 %: 17.6 wt%, $R_w$ = 150 %: 14.8 wt% and $R_w$ = 225 %: 12.2 wt%). In fact, it even exceeds the theoretical 15.5 wt% capacity of structure I methane hydrate, an indication of combined methane uptake by adsorption and hydrate formation. Oversaturating the material beyond $R_w$ = 150 % results in a lower water-to-hydrate conversion as can be seen from the downshifted isotherms for $R_w$ = 225 % (79 % conversion at 7.0 MPa) (Fig. 1b) and $R_w$ = 300 % (23 % conversion at 7.0 MPa) (Fig. S7).

### 3.3 Advanced NMR characterization of CH$_4$ hydrate phase

Enclathration of methane molecules in two different clathrate cages was evidenced by $^2$H (Section S8) and $^{13}$C NMR (Fig. 2, horizontal axes), showing resonances at -4.4 ppm ($^{13}CH_4$ in small $5^{12}$ cages) and -6.7 ppm ($^{13}CH_4$ in large $5^{12}6^2$ cages) in a 1:3 small-to-large-cage ratio, characteristic of an sI type clathrate hydrate.[86] The occurrence of sI methane hydrate was also confirmed by X-ray diffraction (Fig. S9). The heteronuclear correlation experiment summarized in Fig. 2, i.e., $^1$H -$^{13}$C CP-HETCOR, enabled the identification of proton species in close contact with the $^{13}CH_4$ carbon. The strongest correlation (orange bands) in this case occurs between the carbon and protons in methane itself (at -1.07 ppm). In addition, the interaction of the $^{13}CH_4$ carbon to residual water protons (at 4.62 ppm) is evident from the cross contours (highlighted by the green bands), which is expected from methane enclathrated by water molecules.

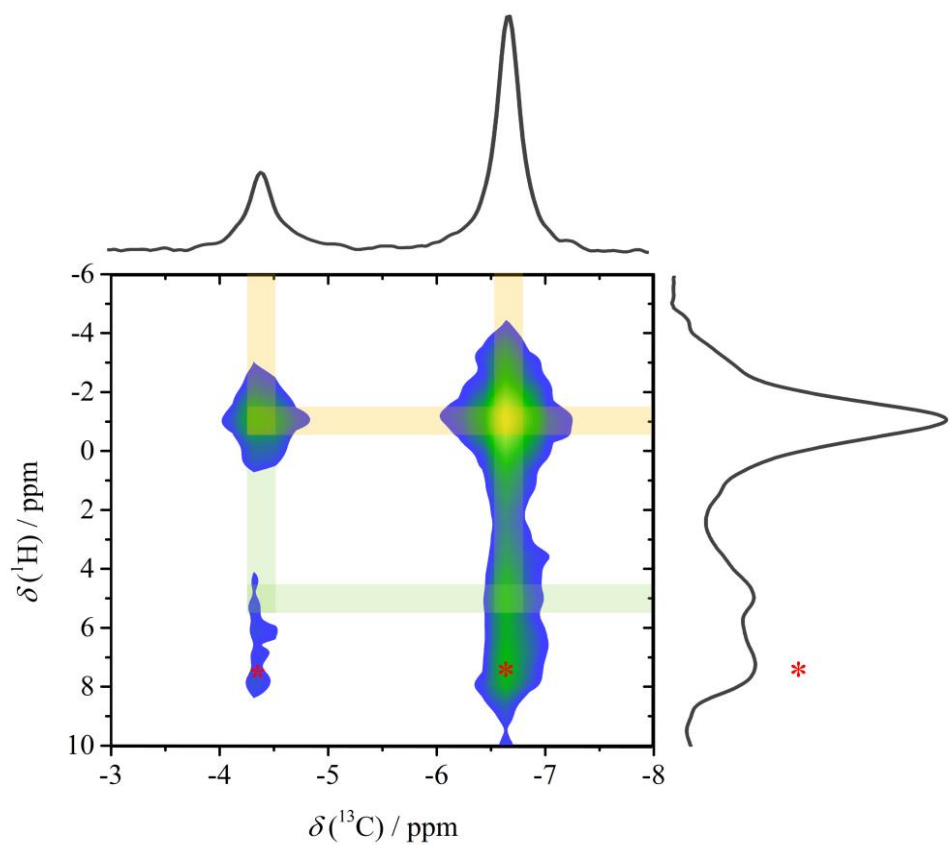

**Fig. 2** 2D $^1$H-$^{13}$C CP-HETCOR spectrum showing the interaction between the carbons and the protons in $^{13}$CH$_4$ itself (orange bands), along with the interaction between two $^{13}$CH$_4$ carbons in a different chemical environment (i.e., in two different clathrate cages) and residual H$_2$O protons (green bands). The experiment was done at a MAS frequency of 4.5 kHz at 223 K. The asterisks represent the spinning sidebands.

The homonuclear $^1$H-$^1$H correlation spectrum of the $^{13}$CH$_4$ hydrate using the RFDR experiment under MAS (4.5 kHz) is shown in Fig. 3. Because of the increased spectral resolution, an additional $^1$H resonance at 2.3 ppm corresponding to the alkyl protons of the C$_8$ group decorating the hydrophobic surface of the material becomes observable.

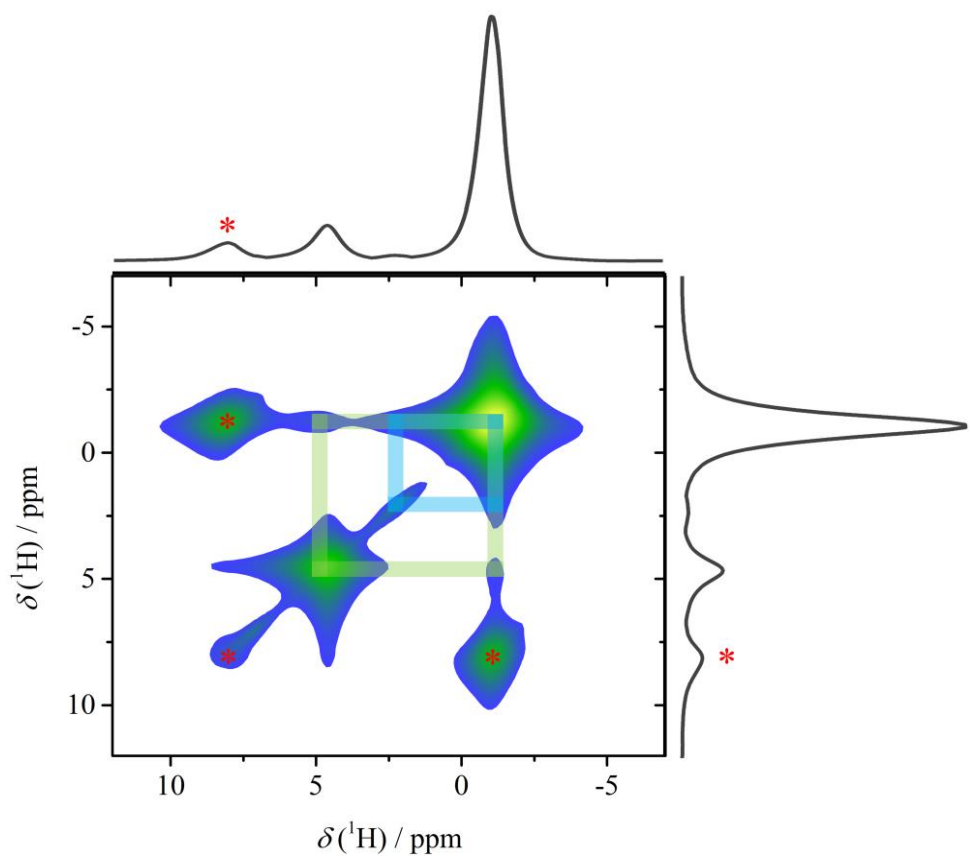

**Fig. 3** 2D $^1$H-$^1$H RFDR correlation spectrum showing the interaction of $^{13}$CH$_4$ and H$_2$O protons (green bands) and a weak dipolar interaction between the hydrophobic surface and $^{13}$CH$_4$ (blue band). The experiment was done at a MAS frequency of 4.5 kHz and a temperature of 223 K. The asterisks represent the spinning sidebands.

Strong dipolar correlation between the $^{13}$CH$_4$ protons and the water protons is evident from the cross-contour intensities marked with green bands. Such an interaction is not plausible when $^{13}$CH$_4$ and water stay in separate, immiscible phases, and points towards the presence of a methane hydrate phase. A weak interaction of the hydrophobic surface protons of the C$_8$ chains and the enclathrated $^{13}$CH$_4$ protons is also present (blue bands) in the system, implying spatial proximity between the two molecules and, by extension, the presence of enclathrated methane in the mesopores of SBA-15 C$_8$. Intuitively, one might also expect a correlation between the alkyl protons and the water protons. However, since D$_2$O was used instead of H$_2$O, and given the small amount of (residual) water molecules in contact with the hydrophobic C$_8$ chains, this correlation is expected to give a signal below the detection limit of the measurement.

## 3.4 Recyclability of hydrate phase

In a next step, the reusability of the octyl-grafted material was evaluated by measuring three successive isotherms on SBA-15 $C_8$ $R_w$ = 150 %, whereby in between cycles an evacuation period of 20 min was adopted (Fig. 4a). All three measurements coincide almost completely over the assessed pressure range. Over consecutive cycles, a small increase in the equilibrium methane uptake at the initial formation pressure (1.5 MPa) can be observed. Nevertheless, the results do not indicate a degradation of the solid material or loss of its hydrate promoting nature over multiple measurements.

More striking, however, is the immensely enhanced hydrate nucleation and growth rate in a second and third isotherm measurement. On Fig. 4b, showing the fractional methane uptake over time at 1.5 MPa for the three isotherms, a sigmoidal uptake curve is observed for the first isotherm measurement, reaching 90 % and 95 % of the equilibrium capacity ($t_{90}$ and $t_{95}$) in 3.5 h (207 min) and 3.8 h (230 min), respectively. However, for both the second and third measurement, no nucleation period is observed and $t_{90}$ and $t_{95}$ decrease to 7.8 and 16.8 min, enhancing the kinetics by a factor of 26.5 and 13.7, respectively (Fig. 4c).

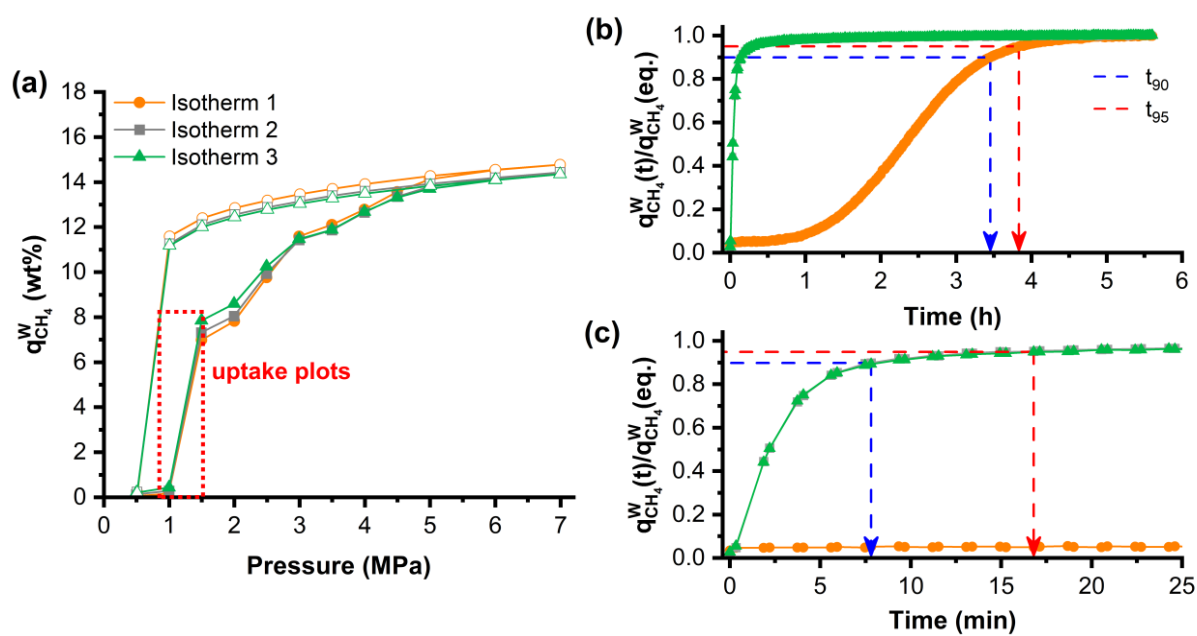

**Fig. 4** (a) Three consecutive methane hydrate isotherm measurements on SBA-15 $C_8$ $R_w$ = 150 % at 248 K, connecting the sample to vacuum for 20 min in between cycles. (b and c) Methane uptake ($q_{CH_4}^w(t)$) normalized to the equilibrium capacity ($q_{CH_4}^w(eq.)$), upon increasing $CH_4$ pressure from 1.0 to 1.5 MPa, as function of time.

To determine whether the strongly enhanced methane hydrate reformation kinetics is a lasting phenomenon, 20 subsequent pressure-driven formation-dissociation cycles were performed on SBA-15 $C_8$ $R_w$ = 150 % at 248 K. Based on the isotherm shape (Fig. 4a), the formation and dissociation pressures were set to 1.5 and 0.5 MPa, respectively. In the first cycle the formation and dissociation times ($t^f$ and $t^d$) were set to 6 and 1 h, respectively, while for the second to twentieth cycle they were set to 2 and 1 h, respectively (Fig. 5a).

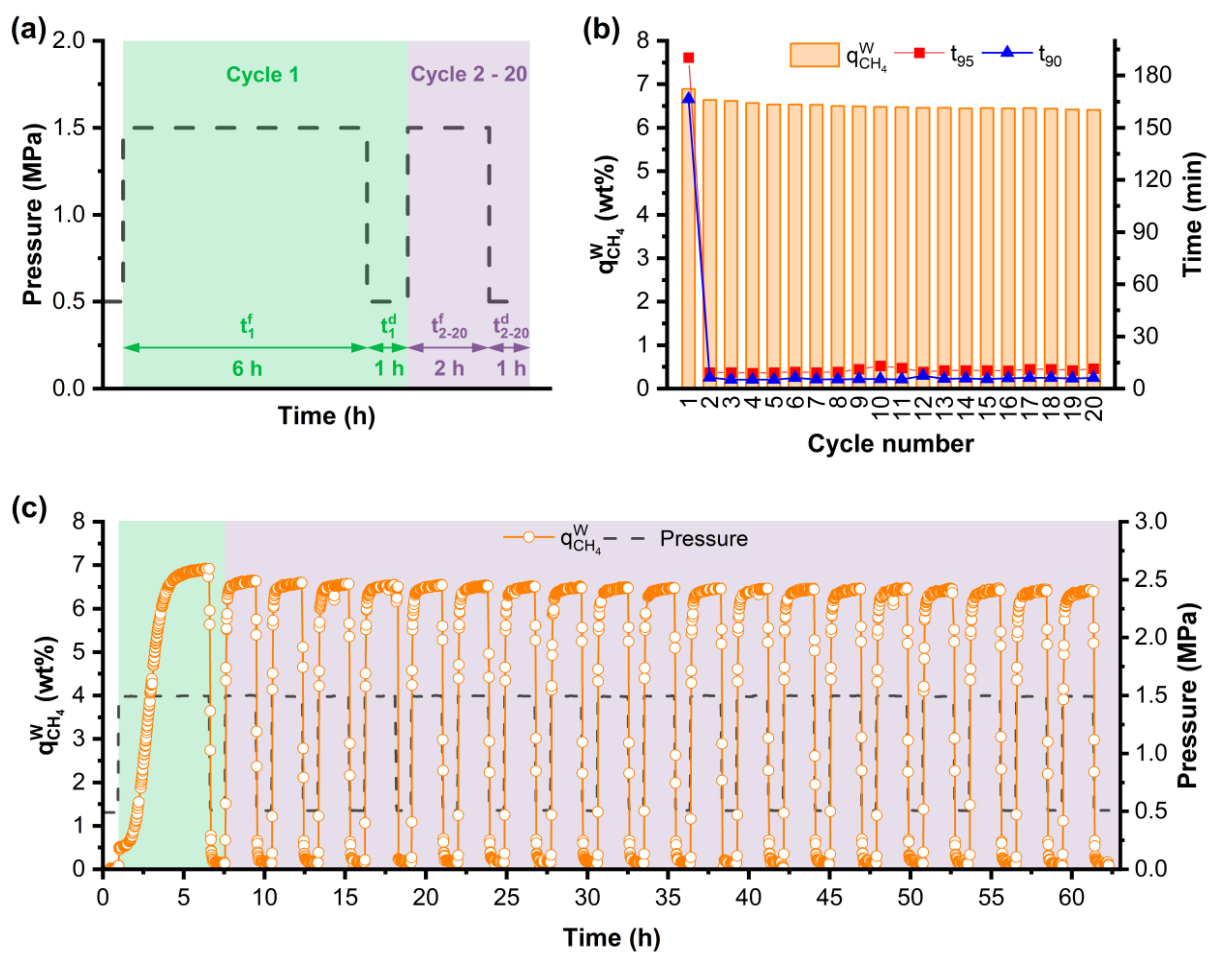

**Fig. 5** Gravimetric uptake-release experiments on SBA-15 $R_w$ = 150 % at 248 K. (a) Programmed pressure profile of the 20 cycles. (b) Hydrate storage capacity, $t_{90}$ and $t_{95}$ of each cycle. (c) Hydrate storage capacity as function of time for the subsequent pressure-induced uptake-release cycles.

Looking at the first cycle (**Fig.** Fig. 5c), a similar sigmoidal uptake curve as in Fig. 4a can be observed, albeit with a slightly lower $t_{90}$ (166.5 min) and $t_{95}$ (190.4 min) due to the larger driving force of $\Delta P$ = 1.0 MPa (as compared to 0.5 MPa in Fig. 4). Dissociation occurs rapidly, as visible from the almost vertical drop in capacity upon lowering pressure ($t_{90}$ = 4.1 min, $t_{95}$ = 7.3 min). In the 19 following formation-dissociation cycles, the enhanced formation kinetics remain (average $t_{90}$ = 5.8 min, $t_{95}$ = 10.4 min), yet a slight decrease in capacity occurs ($q_{CH_4}^W$(cycle 20) = 93 % $q_{CH_4}^W$(cycle 1)) (Fig. 5b,c). For all cycles, there are no clear differences in the dissociation kinetics among the cycles (Fig. S10) and compared to the first cycle. At the end of the depressurization period, capacity has lowered to 0.14 wt% on average. The small amount of methane present in the system is likely to originate from physisorption on non-wetted pore surfaces, or from a small amount of trapped hydrate (0.9 %), as previously shown by Casco and co-workers for methane hydrate formation by benzene-PMO, a hydrophobic mesoporous silica.[49]

A small residual signal attributed to enclathrated methane is indeed observed in the static $^{13}$C NMR spectrum of methane hydrate confined by SBA-15 $C_8$ ($R_w$ = 150 %), formed at 6.0 MPa at 248 K and subsequently depressurized below 0.5 MPa at 248 K (Fig. S11), supporting the trapped hydrate hypothesis. The static $^2$H NMR spectrum of the same sample however shows no trace of the clathrate hydrate structure, most likely due to its limited occurrence. The $^2$H signal comprises a Pake pattern with quadrupole parameters $C_Q$ of 216.5 kHz and $\eta_Q$ of 0.1 A, corresponding to hexagonal ice, next to one broad and one sharp component, with full widths at half maximum (FWHM) of 6.712 kHz and 0.476 kHz, respectively (Fig. 6a). Interestingly, this sharp resonance is not observed for SBA-15 $C_8$ ($R_w$ = 150 %) cooled to 248 K in the absence of methane, i.e., conditions prior to hydrate formation (Fig. 6b). Line narrowing in static NMR experiments can either originate from an increase in mobility, i.e., mobile $D_2O$ molecules in the center of an empty clathrate cage or in a body of confined, hyper-cooled liquid water, or from an increased ordering of the local water structure with respect to its amorphous surroundings, greatly reducing the dipolar broadening induced by chemical shift anisotropy (CSA).[87] The presence of such mobile and/or more structured $D_2O$ phases in combination with residual methane hydrate nuclei, can help rationalize the increased rate at which methane is enclathrated in cycles 2-20 (and in any subsequent cycle for that matter), as it could provide shortcuts for the seemingly slow process of methane dispersion, solubilization and hydrate nucleation and growth from a deeply cooled water matrix comprising amorphous ice and hexagonal $D_2O$ ice.

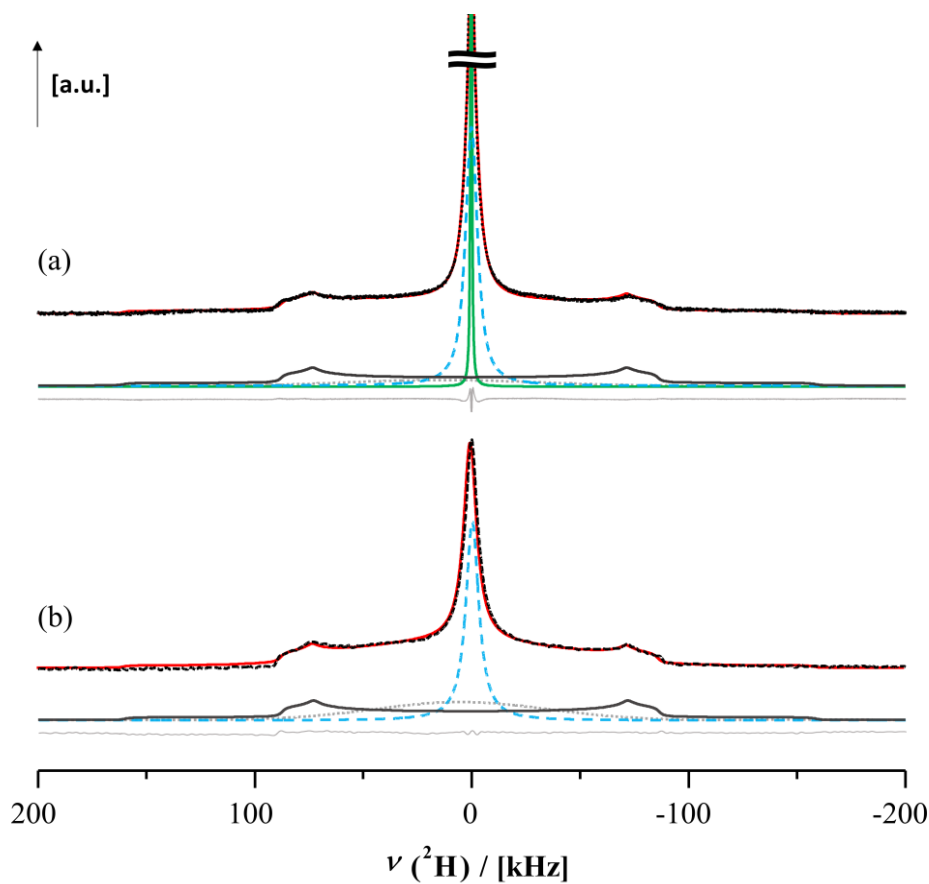

**Fig. 6** $^2$H static NMR spectrum (a) after methane hydrate formation in pre-wet SBA-15 C$_8$ and subsequent methane removal at 0.1 MPa and 248 K, and (b) before exposure of the pre-wet SBA-15 C8 host to methane pressure at 248 K. Using DMfit program, the upper spectrum can be simulated using one double Pake pattern (grey trace) with a quadrupole coupling constant (C$_Q$) of 216.5 kHz and an asymmetry parameter ($h_Q$) of 0.1, one sharp component (green trace, FWHM = 0.476 kHz), one broad component (blue trace, FWHM = 6.712 kHz) and one very broad component (light grey trace, dotted line) assigned to highly immobile deuterons. The simulated spectrum is highlighted in red, the residual trace, i.e., the difference between experiment and simulation, is shown in grey (full line) at the bottom of each decomposition. The lower $^2$H NMR spectrum can be simulated with a similar Pake pattern (grey trace, C$_Q$ = 216.5 kHz and $h_Q$ = 0.1), one broad component (blue trace, FWHM = 7.235 kHz) and one very broad component (light grey trace, dotted line). A full-scale image is provided in Supporting Information Fig. S12.

## Conclusion

This study explored the potential of a hydrophobized SBA-15 material to promote methane hydrate nucleation, growth, and reformation, in terms of equilibrium and kinetics. By grafting octyl groups onto the pore surface, i.e., SBA-15 C$_8$, methane hydrate formation is enhanced as evidenced by the higher methane uptake capacity and a water-to-hydrate conversion of 96 % for SBA-15 $C_8$ with $R_w$ = 150 % compared to 71 % for pristine SBA-15 at 7 MPa and 248 K. Combined evidence from X-ray diffraction and 2D $^1$H-$^{13}$C and $^1$H-$^1$H correlation NMR spectroscopy confirms the presence of methane enclathrated in structure I hydrate, both in interparticle volumes and inside the mesopores of SBA-15 $C_8$. The rate of hydrate formation for SBA-15 $C_8$ ($R_w$ = 150 %) observed during the very first cycle was significantly sped up in subsequent pressure-induced formation-dissociation of the clathrate structure at 248 K, with a reduction of $t_{90}$ from 3.5 h to 7.8 min. This effect was shown to last for at least 19 subsequent cycles. Analysis of the dissociated clathrate hydrate structure with $^{13}$C and $^2$H NMR revealed the presence of a small amount of residual methane hydrate in combination with an amorphous yet highly ordered $D_2O$ ice phase, respectively, which were not present in samples which did not undergo methane hydrate formation. This could help rationalize the enhanced reformation kinetics observed on a macroscopic level. These initial results are very promising for the purpose of developing a hydrate-based methane storage technology. In addition, isothermal operation of the system at 248 K allows to cut energy consumption in subsequent hydrate dissociation-formation cycles by 50% by obviating the need to cool down the system from 293 to 248 K during each formation step (assuming a dissociation enthalpy of -45 kJ/mol[88]). Design of porous materials promoting hydrate (re)formation at higher $R_w$ values is currently being explored in order to maximize the overall storage capacity of the system.

## Conflicts of interest

There are no conflicts to declare

## Acknowledgements

**To be completed by co-authors**

M.H. acknowledges FWO for an FWO-SB fellowship. All authors acknowledge VLAIO for Moonshot funding (ARCLATH, n° HBC.2019.0110, ARCLATH2, n° HBC.2021.0254). J.A.M. acknowledges the Flemish Government for long-term structural funding (Methusalem) and department EWI for infrastructure investment via the Hermes Fund (AH.2016.134). NMRCoRe thanks the Flemish government for financial support as International Research Infrastructure (I001321N: Nuclear Magnetic Resonance Spectroscopy Platform for Molecular Water Research) and acknowledges infrastructure funding from the Department EWI via the Hermes Fund (AH.2016.134). J.A.M. acknowledges the European Research Council (ERC) for an Advanced Research Grant under the European Union's Horizon 2020 research and innovation program under grant agreement No. 834134 (WATUSO).

## Data availability

# References

1	E. D. Sloan and C. A. Koh, *Clathrate Hydrates of Natural Gases*, CRC Press, Boca Raton, 3th editio., 2008.

2	J. Klauda and S. Sandler, *Energy Fuels - ENERG FUEL*, , DOI:10.1021/ef049798o.

3	C. Ruppel, *J. Chem. Eng. Data*, 2015, **60**, 429–436.

4	K. A. Kvenvolden, *Chem. Geol.*, 1988, **71**, 41–51.

5	C. A. Koh, *Chem. Soc. Rev.*, 2002, **31**, 157–167.

6	H. P. Veluswamy, A. Kumar, Y. Seo, J. D. Lee and P. Linga, *Appl. Energy*, 2018, 216, 262–285.

7	Z. R. Chong, S. H. B. Yang, P. Babu, P. Linga and X. Sen Li, *Appl. Energy*, 2016, **162**, 1633–1652.

8	Y. Zhang, J. Zhao, G. Bhattacharjee, H. Xu, M. Yang, R. Kumar and P. Linga, *Energy Environ. Sci.*, 2022, **15**, 5362–5378.

9	A. Kumar, O. S. Kushwaha, P. Rangsunvigit, P. Linga and R. Kumar, *Can. J. Chem. Eng.*, 2016, **94**, 2160–2167.

10	C. Taylor, K. Miller, C. Koh and E. Sloan, *Chem. Eng. Sci.*, 2007, **62**, 6524–6533.

11	A. Y. Manakov, N. V Penkov, T. V Rodionova, A. N. Nesterov and E. E. Fesenko Jr, *Russ. Chem. Rev.*, 2017, **86**, 845–869.

12	E. Chaturvedi, S. Laik and A. Mandal, *Chinese J. Chem. Eng.*, 2021, **32**, 1–16.

13	H. P. Veluswamy, S. Kumar, R. Kumar, P. Rangsunvigit and P. Linga, *Fuel*, 2016, **182**, 907–919.

14	T. Altamash, J. M. S. S. Esperança and M. Tariq, *Physchem*, 2021, 1, 272–287.

15	M. J. Shultz and T. H. Vu, *J. Phys. Chem. B*, 2015, **119**, 9167–9172.

16	H. P. Veluswamy, A. J. H. Wong, P. Babu, R. Kumar, S. Kulprathipanja, P. Rangsunvigit and P. Linga, *Chem. Eng. J.*, 2016, **290**, 161–173.

17	L. Borchardt, M. E. Casco and J. Silvestre-Albero, *ChemPhysChem*, 2018, 19, 1298–1314.

18	Y. Qin, L. Shang, Z. Lv, J. He, X. Yang and Z. Zhang, *J. Energy Chem.*, 2022, **74**, 454–480.

19	P. Warrier, M. N. Khan, V. Srivastava, C. M. Maupin and C. A. Koh, *J. Chem. Phys.*, 2016, **145**, 211705.

20	Y. Long, J. C. Palmer, B. Coasne, M. Śliwinska-Bartkowiak and K. E. Gubbins, *Phys. Chem. Chem. Phys.*, 2011, **13**, 17163–17170.

21	Y. (龙云) Long, J. C. Palmer, B. Coasne, M. Śliwinska-Bartkowiak, G. Jackson, E. A. Müller and K. E. Gubbins, *J. Chem. Phys.*, 2013, **139**, 144701.

22   M. Śliwińska-Bartkowiak, H. Drozdowski, M. Kempiński, M. Jażdżewska, Y. Long, J. C. Palmer and K. E. Gubbins, *Phys. Chem. Chem. Phys.*, 2012, **14**, 7145–7153.

23   M. E. Casco, J. Silvestre-Albero, A. J. Ramírez-Cuesta, F. Rey, J. L. Jordá, A. Bansode, A. Urakawa, I. Peral, M. Martínez-Escandell, K. Kaneko and F. Rodríguez-Reinoso, *Nat. Commun.*, 2015, **6**, 1–8.

24   M. E. Casco, J. L. Jordá, F. Rey, F. Fauth, M. Martinez-Escandell, F. Rodríguez-Reinoso, E. V. Ramos-Fernández and J. Silvestre-Albero, *Chem. - A Eur. J.*, 2016, **22**, 10028–10035.

25   L. Wang, M. Dou, Y. Wang, Y. Xu, Y. Li, Y. Chen and L. Li, *ACS Omega*, 2022, **7**, 33666–33679.

26   X. Liu, L. Zhou, J. Li, Y. Sun, W. Su and Y. Zhou, *Carbon N. Y.*, 2006, **44**, 1386–1392.

27   J. Liu, Y. Zhou, Y. Sun, W. Su and L. Zhou, *Carbon N. Y.*, 2011, **49**, 3731–3736.

28   M. E. Casco, C. Cuadrado-Collados, M. Martínez-Escandell, F. Rodríguez-Reinoso and J. Silvestre-Albero, *Carbon N. Y.*, 2017, **123**, 299–301.

29   L. Borchardt, W. Nickel, M. Casco, I. Senkovska, V. Bon, D. Wallacher, N. Grimm, S. Krause and J. Silvestre-Albero, *Phys. Chem. Chem. Phys.*, 2016, **18**, 20607–20614.

30   M. E. Casco, E. Zhang, S. Grätz, S. Krause, V. Bon, D. Wallacher, N. Grimm, D. M. Többens, T. Hauß and L. Borchardt, *J. Phys. Chem. C*, 2019, **123**, 24071–24079.

31   L. Zhou, Y. Sun and Y. Zhou, *AIChE J.*, 2002, **48**, 2412–2416.

32   A. Celzard and J. F. Marêché, *Fuel*, 2006, **85**, 957–966.

33   A. Perrin, A. Celzard, J. F. Marêché and G. Furdin, *Energy and Fuels*, 2003, **17**, 1283–1291.

34   M. J. D. Mahboub, A. Ahmadpour and H. Rashidi, *Ranliao Huaxue Xuebao/Journal Fuel Chem. Technol.*, 2012, **40**, 385–389.

35   M. E. Casco, F. Rey, J. L. Jordá, S. Rudić, F. Fauth, M. Martínez-Escandell, F. Rodríguez-Reinoso, E. V. Ramos-Fernández and J. Silvestre-Albero, *Chem. Sci.*, 2016, **7**, 3658–3666.

36   L. Mu, B. Liu, H. Liu, Y. Yang, C. Sun and G. Chen, *J. Mater. Chem.*, 2012, **22**, 12246–12252.

37   S. Denning, A. A. A. Majid, J. M. Lucero, J. M. Crawford, M. A. Carreon and C. A. Koh, *ACS Appl. Mater. Interfaces*, 2020, **12**, 53510–53518.

38   C. Cuadrado-Collados, G. Mouchaham, L. Daemen, Y. Cheng, A. Ramirez-Cuesta, H. Aggarwal, A. Missyul, M. Eddaoudi, Y. Belmabkhout and J. Silvestre-Albero, *J. Am. Chem. Soc.*, 2020, **142**, 13391–13397.

39   D. Kim, Y. H. Ahn and H. Lee, *J. Chem. Eng. Data*, 2015, **60**, 2178–2185.

40   C. Chen, Y. Li and J. Cao, *Catalysts*, 2022, **12**, 1–13.

41   E. Andres-Garcia, A. Dikhtiarenko, F. Fauth, J. Silvestre-Albero, E. V. Ramos-Fernández, J. Gascon, A. Corma and F. Kapteijn, *Chem. Eng. J.*, 2019, **360**, 569–576.

42  X. Zang, J. Du, D. Liang, S. Fan and C. Tang, *Chinese J. Chem. Eng.*, 2009, **17**, 854–859.

43  D. L. Zhong, Z. Li, Y. Y. Lu, J. Le Wang, J. Yan and S. L. Qing, *Ind. Eng. Chem. Res.*, 2016, **55**, 7973–7980.

44  Y. Zhao, J. Zhao, W. Liang, Q. Gao and D. Yang, *Fuel*, 2018, **220**, 185–191.

45  U. Karaaslan and M. Parlaktuna, *Energy and Fuels*, 2002, **16**, 1413–1416.

46  M. T. Sun, F. P. Song, G. D. Zhang, J. Z. Li and F. Wang, *Fuel*, 2021, **288**, 119676.

47  P. Linga, C. Haligva, S. C. Nam, J. A. Ripmeester and P. Englezos, *Energy and Fuels*, 2009, **23**, 5496–5507.

48  T. Uchida, T. Ebinuma and T. Ishizaki, *J. Phys. Chem. B*, 1999, **103**, 3659–3662.

49  M. E. Casco, S. Grätz, D. Wallacher, N. Grimm, D. M. Többens, M. Bilo, N. Speil, M. Fröba and L. Borchardt, *Chem. Eng. J.*, 2021, **405**, 126955.

50  X. Liu, D. Liu, W. Xie, X. Cui and Y. Chen, *J. Chem. Eng. Data*, 2018, **63**, 1767–1772.

51  D. L. Zhong, S. Y. He, D. J. Sun and C. Yang, *Energy Procedia*, 2014, **61**, 1573–1576.

52  B. O. Carter, W. Wang, D. J. Adams and A. I. Cooper, *Langmuir*, 2010, **26**, 3186–3193.

53  W. Wang, C. L. Bray, D. J. Adams and A. I. Cooper, *J. Am. Chem. Soc.*, 2008, **130**, 11608–11609.

54  J. Park, K. Shin, J. Kim, H. Lee, Y. Seo, N. Maeda, W. Tian and C. D. Wood, *J. Phys. Chem. C*, 2015, **119**, 1690–1699.

55  S. Fan, L. Yang, Y. Wang, X. Lang, Y. Wen and X. Lou, *Chem. Eng. Sci.*, 2014, **106**, 53–59.

56  J. Wang, R. Wang, R. H. Yoon and Y. Seol, *J. Chem. Eng. Data*, 2015, **60**, 383–388.

57  H. Li and L. Wang, *Fuel*, 2015, **140**, 440–445.

58  F. Filarsky, C. Schmuck and H. J. Schultz, *Ind. Eng. Chem. Res.*, 2019, **58**, 16687–16695.

59  Z. R. Chong, M. Yang, B. C. Khoo and P. Linga, *Ind. Eng. Chem. Res.*, 2016, **55**, 7981–7991.

60  V. C. Nair, S. Ramesh, G. A. Ramadass and J. S. Sangwai, *J. Pet. Sci. Eng.*, 2016, **147**, 547–559.

61  Y. P. Handa and D. Stupin, *J. Phys. Chem.*, 1992, **96**, 8599–8603.

62  Y. Seo, H. Lee and T. Uchida, *Langmuir*, 2002, **18**, 9164–9170.

63  K. Inkong, H. P. Veluswamy, P. Rangsunvigit, S. Kulprathipanja and P. Linga, *Ind. Eng. Chem. Res.*, 2019, **58**, 22178–22192.

64  D. H. Smith, J. W. Wilder and K. Seshadri, *AIChE J.*, 2002, **48**, 393–400.

65  L. Zhou, X. Liu, Y. Sun, J. Li and Y. Zhou, *J. Phys. Chem. B*, 2005, **109**, 22710–

22714.

66   S. Denning, A. A. A. Majid and C. A. Koh, *J. Phys. Chem. C*, 2022, **126**, 11800–11809.

67   J.-L. Chen, P. Xiao, D.-X. Zhang, G.-J. Chen, C.-Y. Sun, Q.-L. Ma, M.-K. Yang and E.-B. Zou, *Front. Chem.*, , DOI:10.3389/fchem.2020.00294.

68   N. N. Nguyen, A. V. Nguyen, K. M. Steel, L. X. Dang and M. Galib, *J. Phys. Chem. C*, 2017, **121**, 3830–3840.

69   N. N. Nguyen, M. Galib and A. V. Nguyen, *Energy and Fuels*, 2020, **34**, 6751–6760.

70   D. T. Bowron, A. Filipponi, M. A. Roberts and J. L. Finney, *Phys. Rev. Lett.*, 1998, **81**, 4164–4167.

71   A. Tinti, A. Giacomello, Y. Grosu and C. M. Casciola, *Proc. Natl. Acad. Sci.*, 2017, **114**, E10266–E10273.

72   A. Cavazzini, N. Marchetti, L. Pasti, R. Greco, F. Dondi, A. Laganà, A. Ciogli and F. Gasparrini, *Anal. Chem.*, 2013, **85**, 19–22.

73   H. Moon, R. P. Collanton, J. I. Monroe, T. M. Casey, M. S. Shell, S. Han and S. L. Scott, *J. Am. Chem. Soc.*, 2022, **144**, 1766–1777.

74   S. Eslami, B. Farhangdoost, H. Shahverdi and M. Mohammadi, *Greenh. Gases Sci. Technol.*, 2021, **11**, 939–953.

75   M. Borsboom, W. Bras, I. Cerjak, D. Detollenaere, D. Glastra van Loon, P. Goedtkindt, M. Konijnenburg, P. Lassing, Y. K. Levine, B. Munneke, M. Oversluizen, R. van Tol and E. Vlieg, *J. Synchrotron Radiat.*, 1998, **5**, 518–520.

76   M. Houlleberghs, J. A. Martens and E. Breynaert, *J. Synchrotron Radiat.*, 2018, **25**, 1893–1894.

77   V. Meynen, P. Cool and E. F. Vansant, *Microporous Mesoporous Mater.*, 2009, **125**, 170–223.

78   T. Yasmin and K. Müller, *J. Chromatogr. A*, 2011, **1218**, 6464–6475.

79   C. V. Chandran and T. Bräuniger, *J. Magn. Reson.*, 2009, **200**, 226–232.

80   D. Massiot, F. Fayon, M. Capron, I. King, S. Le Calvé, B. Alonso, J.-O. Durand, B. Bujoli, Z. Gan and G. Hoatson, *Magn. Reson. Chem.*, 2002, **40**, 70–76.

81   T. Park, J. Y. Lee and T. H. Kwon, *Energy and Fuels*, 2018, **32**, 5321–5330.

82   L. Borchardt, W. Nickel, M. Casco, I. Senkovska, V. Bon, D. Wallacher, N. Grimm, S. Krause and J. Silvestre-Albero, *Phys. Chem. Chem. Phys.*, 2016, **18**, 20607–20614.

83   P. Mileo, S. M. J. Rogge, M. Houlleberghs, E. Breynaert, J. Martens and V. Van Speybroeck, *J. Mater. Chem. A*, , DOI:10.1039/d1ta03105h.

84   K. Kroenlein, C. D. Muzny, A. Kazakov, V. V. Diky, R. D. Chirico, E. D. Sloan and M. Frenkel, Clathrate Hydrate Physical Property Database, http://gashydrates.nist.gov/.

85   J. Wang, R. Wang, R. H. Yoon and Y. Seol, *J. Chem. Eng. Data*, 2015, **60**, 383–388.

86   Y.-T. Seo and H. Lee, *Korean J. Chem. Eng.*, 2003, **20**, 1085–1091.

87  Y. J. Lee, T. Murakhtina, D. Sebastiani and H. W. Spiess, *J. Am. Chem. Soc.*, 2007, **129**, 12406–12407.

88  H. Liang, Y. Duan, J. Pei and N. Wei, *Front. Energy Res.*, , DOI:10.3389/fenrg.2021.743296.